\documentclass[10pt,a4paper]{article}
\usepackage{amsmath}
\usepackage{amssymb}
\usepackage{amsfonts}
\usepackage{amstext}
\usepackage{amsbsy}
\usepackage[mathscr]{eucal}
\usepackage{graphicx}
\usepackage{color}
\newcommand{\beq}{\begin{equation}}
\newcommand{\eeq}{\end{equation}}
\newcommand{\beqa}{\begin{eqnarray}}
\newcommand{\eeqa}{\end{eqnarray}}
\newcommand{\ket}[1]{| #1 \rangle}

\title{\Large\textbf{Combinatorial structure of a holonomic controlled phase gate}}

\author{\textit{ Hoshang Heydari}\\
        \small\textit{Physics Department, Stockholm University 10691 Stockholm Sweden}\\
\\\small\textit{Email: hoshang@fysik.su.se}}
\begin{document}
\maketitle

\begin{abstract}
We investigate the combinatorial structures of a holonomic controlled quantum gate based on  toric varieties. In particular, we in detail   discuss the combinatorial structures of a two-qubit holonomic controlled quantum gate on a two-qubit, a three-qubit, and a four-qubit quantum states. Our results show interesting relations between toric varieties of qubit systems and action of a two-qubit holonomic quantum gate on multi-qubit states.
\end{abstract}


\maketitle

\section{Introduction}

Holonomic quantum computing is a model of computing which is  based on geometrical prosperities of quantum systems and could e.g., result in a less error prone computational model. In the holonomic model
a quantum gate is constructed from the geometric phase.
The Hamiltonian is varied adiabatically through a closed curve in a  parameter space.
In this paper we are interested to visualized the
action of a holonomic gate on product states  based on  toric varieties. A toric variety
is an irreducible variety $\mathbf{X}$ that satisfies the following conditions. First of all $T=(\mathbf{C}^{*})^{n}$ is a Zariski open subset of $\mathbf{X}$ and the action of
$T$ on itself can extend to an action of  $T$  on the variety
$\mathbf{X}$. We are interested in combinatorial structures of toric
varieties which are discussed in \cite{Hosh5,Ewald}.
Next,  we will consider the following  Hamiltonian of an interacting two qubits spin system
\cite{Ekert}
\begin{eqnarray}
 \nonumber H&=&
H_{0}+H_{int}\\\nonumber&=&
\hbar\omega_{i}S_{iz}\otimes \mathbf{I}_{j}+\hbar\omega_{j}\mathbf{I}_{i}\otimes S_{jz}+2\pi\hbar J S_{iz}\otimes S_{jz}\\\nonumber&=&
\frac{\hbar}{2}\mathrm{diag}(\omega_{i}+\omega_{j}+\pi J,
\omega_{i}-\omega_{j}-\pi J,
-\omega_{i}+\omega_{j}-\pi J,
-\omega_{i}-\omega_{j}+\pi J),
\end{eqnarray}
where $\mathrm{diag}$ denotes a diagonal matrix, $\omega_{i}>\omega_{j}$ are  the transition angular frequency of the two spins, $S_{rz}=\sigma_{iz}/2$ is the Pauli matrix, and $J$ is the coupling constant. In the last part of the above equation the Hamiltonian is written in the basis $\{\ket{\uparrow\uparrow},
\ket{\uparrow\downarrow},
\ket{\downarrow\uparrow},
\ket{\downarrow\downarrow}\}.$ We assume that when the spin $S_{j}$ is in state $\ket{\uparrow}$, the transition frequency of spin $S_{i}$ is $\omega_{+}=\omega_{i}+\pi J$ and when the spin $S_{j}$ is in state $\ket{\downarrow}$, the transition frequency of spin $S_{i}$ is $\omega_{-}=\omega_{i}-\pi J$. Next we define a
 Berry phase shift
\begin{eqnarray}
 \nonumber \gamma_{+}+\gamma_{-}&=&
\pi\left(\cos\theta_{+}-\cos\theta_{-}\right)
=
\pi\left(
\frac{\omega_{+}-\omega}{\sqrt{(\omega_{+}-\omega)^{2}+\omega^{2}_{1}}}-
\frac{\omega_{-}-\omega}{\sqrt{(\omega_{-}-\omega)^{2}+\omega^{2}_{1}}}\right),
\end{eqnarray}
where $\gamma_{\pm}=\mp\pi(1-\cos\theta_{\pm})$ are the Berry phases acquired by spin $S_{i}$ when the spin $S_{j}$ is in the state $\ket{\downarrow}$ and $\ket{\uparrow}$. This phase shift gives the following transformation that is equivalent to the controlled phase gate
\begin{eqnarray}
  U_{CPhase} &=&\left(
                   \begin{array}{cccc}
                     e^{2i(\gamma_{+}+\gamma_{-})} & 0 & 0 & 0 \\
                     0 & e^{-2i(\gamma_{+}+\gamma_{-})} & 0 & 0\\
                    0  & 0 & e^{-2i(\gamma_{+}+\gamma_{-})} & 0 \\
                    0 & 0& 0 & e^{2i(\gamma_{+}+\gamma_{-})} \\
                   \end{array}
                 \right)
\end{eqnarray}
In the following sections we will use this holonomic quantum gate to
 investigate the geometrical structures quantum computing based on toric varieties.

\section{Toric variety}

Let $S\subset \mathbf{R}^{n}$ be finite subset, then a convex polyhedral cone is defined by
$
 \sigma=\mathrm{Cone}(S
 )=\left\{\sum_{v\in S}\lambda_{v}v|\lambda_{v}\geq0\right\}.
$
In this case $\sigma$ is generated by $S$.  In a similar way we define  a polytope by
$
 P=\mathrm{Conv}(S)=\left\{\sum_{v\in S}\lambda_{v}v|\lambda_{v}\geq0, \sum_{v\in S}\lambda_{v}=1\right\}.
$
We also could say that $P$ is convex hull of $S$. A convex polyhedral cone is called simplicial if it is generated by linearly independent set. Now, let $\sigma\subset \mathbf{R}^{n}$ be a convex polyhedral cone and $\langle u,v\rangle$ be a natural pairing between $u\in \mathbf{R}^{n}$ and $v\in\mathbf{R}^{n}$. Then, the dual cone of the $\sigma$ is define by
$$
 \sigma^{\wedge}=\left\{u\in \mathbf{R}^{n*}|\langle u,v\rangle\geq0~\forall~v\in\sigma\right\},
$$
where $\mathbf{R}^{n*}$ is dual of $\mathbf{R}^{n}$.
We call a convex polyhedral cone strongly convex if $\sigma\cap(-\sigma)=\{0\}$.

The algebra of Laurent polynomials is defined by
$
\mathbf{C}[z,z^{-1}]=\mathbf{C}[z_{1},z^{-1}_{1},\ldots,$ $z_{n},z^{-1}_{n}],
$
where $z_{i}=\chi^{e^{*}_{i}}$. The terms  of the form $\lambda \cdot z^{\beta}=\lambda z^{\beta_{1}}_{1}z^{\beta_{2}}_{2}\cdots z^{\beta_{n}}_{n}$ for $\beta=(\beta_{1},\beta_{2},\ldots,\beta_{n})\in \mathbf{Z}$ and $\lambda\in \mathbf{C}^{*}$ are called Laurent monomials. A ring $R$ of Laurent polynomials is called a monomial algebra if it is a $\mathbf{C}$-algebra generated by Laurent monomials. Moreover, for a lattice cone $\sigma$, the ring
$R_{\sigma}=\{f\in \mathbf{C}[z,z^{-1}]:\mathrm{supp}(f)\subset \sigma\}
$
is a finitely generated monomial algebra, where the support of a Laurent polynomial $f=\sum\lambda_{i}z^{i}$ is defined by
$\mathrm{supp}(f)=\{i\in \mathbf
{Z}^{n}:\lambda_{i}\neq0\}.$

 Now, for a lattice cone $\sigma$ we can define an affine toric variety to be the maximal spectrum $\mathbf{X}_{\sigma}=\mathrm{Spec}R_{\sigma}.$
   A toric variety
$\mathbf{X}_{\Sigma}$ associated to a fan $\Sigma$ is the result of gluing affine varieties
$\mathbf{X}_{\sigma}=\mathrm{Spec}R_{\sigma}$ for all $\sigma\in \Sigma$  by identifying $\mathbf{X}_{\sigma}$ with the corresponding Zariski open subset in $\mathbf{X}_{\sigma^{'}}$ if
$\sigma$ is a face of $\sigma^{'}$. That is,
first we take the disjoint union of all affine toric varieties $\mathbf{X}_{\sigma}$ corresponding to the cones of $\Sigma$.
Then by gluing all these affine toric varieties together we get $\mathbf{X}_{\Sigma}$.

\section{Toric variety and holonomic quantum computation}
\begin{figure}\label{Figure 1}
\begin{center}
\includegraphics[scale=0.40]{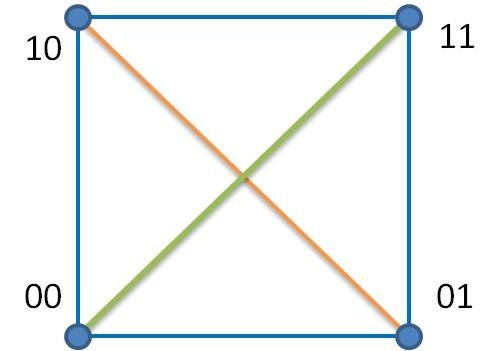}
\end{center}
\caption{Action of the holonomic controlled quantum gate on toric variety of a two-qubit state.}
\end{figure}

\subsection{Two qubit systems}
First we consider a pairs of qubits  $\ket{\Psi}=\sum^{1}_{x_{2}=0}\sum^{1}_{x_{1}=0}
\alpha_{x_{1}x_{2}} \ket{x_{1}x_{2}}$. For this two qubit state the separable state is given by the Segre embedding of
$\mathbf{CP}^{1}\times\mathbf{CP}^{1}=
\{((\alpha^{1}_{0},\alpha^{1}_{1}),(\alpha^{2}_{0},\alpha^{2}_{1})): (\alpha^{1}_{0},\alpha^{1}_{1})\neq0,~(\alpha^{2}_{0},\alpha^{2}_{1})\neq0\}$. Let $z_{1}=\alpha^{1}_{1}(\alpha^{1}_{0})^{-1}$ and $z_{2}=\alpha^{2}_{1}(\alpha^{2}_{0})^{-1}$. Then we can cover $\mathbf{CP}^{1}\times\mathbf{CP}^{1}$ by four charts
\begin{equation}\nonumber
\mathbf{X}_{\check{\sigma}_{1}}=\{(z_{1},z_{2})\},
~\mathbf{X}_{\check{\sigma}_{2}}=\{(z^{-1}_{1},z_{2})\},
~
\mathbf{X}_{\check{\sigma}_{3}}=\{(z_{1},z^{-1}_{2})\},~
\mathbf{X}_{\check{\sigma}_{4}}=\{(z^{-1}_{1},z^{-1}_{2})\}.
\end{equation}
 The fan $\Sigma$ for $\mathbf{CP}^{1}\times\mathbf{CP}^{1}$ has edges spanned by
 $(0,0),(0,1),(1,0),(1,1)$.
Now  we consider the following holonomic controlled quantum gate
  \begin{eqnarray}\label{HQG}
  U_{CPhase} &=&\left(
                   \begin{array}{cccc}
                     e^{i\varphi_{1}} & 0 & 0 & 0 \\
                     0 & e^{i\varphi_{2}} & 0 & 0\\
                    0  & 0 & e^{i\varphi_{2}} & 0 \\
                    0 & 0& 0 & e^{i\varphi_{1}} \\
                   \end{array}
                 \right),
\end{eqnarray}
where  $\varphi_{2}=-\varphi_{1}=-2(\gamma_{+}+\gamma_{-})$. If we apply this gate to  a pure two-qubit product state
$\ket{\Psi}=H\otimes H\ket{00}=\frac{1}{2}(\ket{00}+
\ket{01}+\ket{10}+\ket{11})$,
 then we get
\begin{eqnarray}
U_{CPhase}\ket{\Psi}&=&
\frac{1}{2}(e^{i\varphi_{1}} (\ket{00}+\ket{11})+
e^{i\varphi_{2}} (\ket{01}+ \ket{10} ).
\end{eqnarray}
In this case we can see directly that the phase factors $e^{i\varphi_{1}}$ and $e^{i\varphi_{2}}$
 correspond to diagonal lines in the toric variety of a given two-qubit state. Thus the action of
 the holonomic controlled quantum gate can be seen from a toric variety, see Figure 1.  We also can calculate the concurrence of the evolved state as follows
\begin{eqnarray}
C(U_{CPhase}\ket{\Psi})&=&2|\alpha_{00}\alpha_{11}-\alpha_{01}\alpha_{10}|=
2|\frac{1}{4}e^{i2\varphi_{1}} -\frac{1}{4}
e^{i2\varphi_{2}} |
=|\sin2\varphi_{1}|.
\end{eqnarray}

\subsection{Three qubit systems}
Next, we will discuss a three-qubit state $\ket{\Psi}=\sum^{1}_{x_{3},x_{2},x_{1}=0}
\alpha_{x_{1}x_{2}x_{3}} \ket{x_{1}x_{2}x_{3}}$.
For this  state the separable state is given by the Segre embedding of $\mathbf{CP}^{1}\times\mathbf{CP}^{1}\times\mathbf{CP}^{1}=
\{((\alpha^{1}_{0},\alpha^{1}_{1}),(\alpha^{2}_{0},\alpha^{2}_{1}),(\alpha^{3}_{0},\alpha^{3}_{1}))): (\alpha^{1}_{0},\alpha^{1}_{1})\neq0,~(\alpha^{2}_{0},\alpha^{2}_{1})\neq0
,~(\alpha^{3}_{0},\alpha^{3}_{1})\neq0\}$.
Now, for example, let $z_{1}=\alpha^{1}_{1}/\alpha^{1}_{0}$,
$z_{2}=\alpha^{2}_{1}/\alpha^{2}_{0}$, and $z_{3}=\alpha^{3}_{1}/\alpha^{3}_{0}$.
  Then we can cover $\mathbf{CP}^{1}\times\mathbf{CP}^{1}\times\mathbf{CP}^{1}$ by eight charts
\begin{eqnarray}\nonumber
\nonumber &&
\mathbf{X}_{\check{\Delta}_{1}}=\{(z_{1},z_{2},z_{3})\},
~\mathbf{X}_{\check{\Delta}_{2}}=\{(z^{-1}_{1},z_{2},z_{3})\},
~\mathbf{X}_{\check{\Delta}_{3}}=\{(z_{1},z^{-1}_{2},z_{3})\},~\\\nonumber&&
\mathbf{X}_{\check{\Delta}_{4}}=\{(z_{1},z_{2},z^{-1}_{3})\},
\mathbf{X}_{\check{\Delta}_{5}}=\{(z^{-1}_{1},z^{-1}_{2},z_{3})\},
~\mathbf{X}_{\check{\Delta}_{6}}=\{(z^{-1}_{1},z_{2},z^{-1}_{3})\},~\\\nonumber&&
\mathbf{X}_{\check{\Delta}_{7}}=\{(z_{1},z^{-1}_{2},z^{-1}_{3})\},~
\mathbf
{X}_{\check{\Delta}_{8}}=\{(z^{-1}_{1},z^{-1}_{2},z^{-1}_{3})\}.
\end{eqnarray}
\begin{figure}\label{Figure 2}
\begin{center}
\includegraphics[scale=0.40]{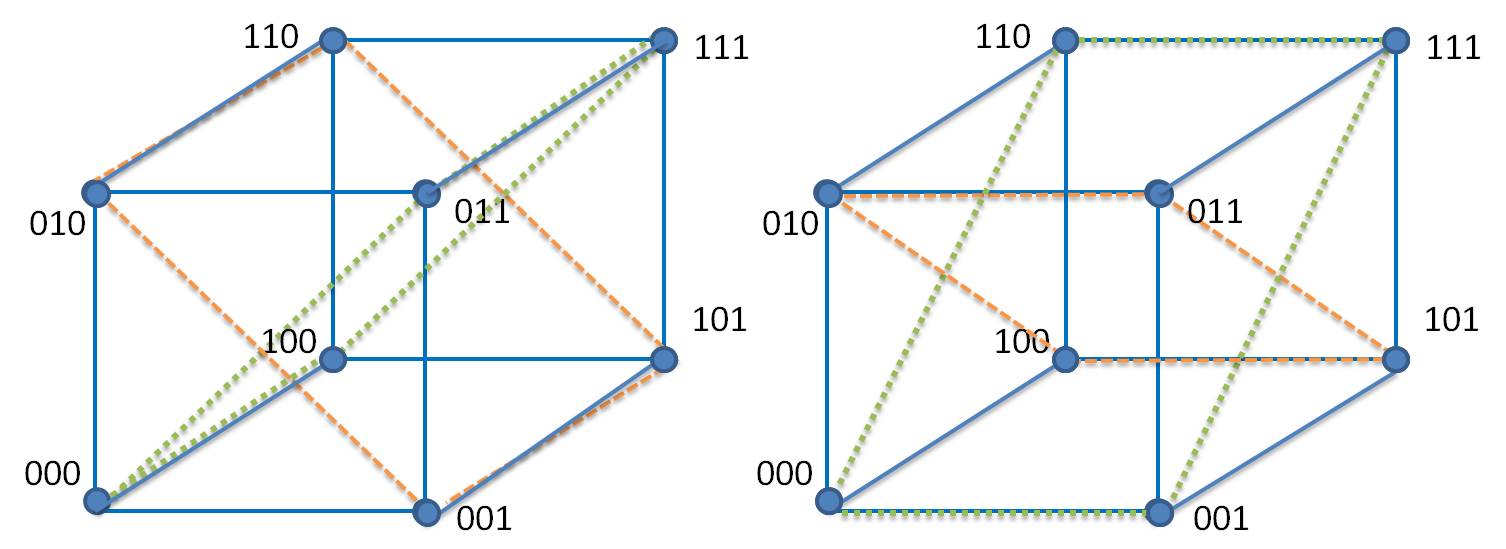}
\end{center}
\caption{Action of the holonomic controlled quantum gate on toric variety of a three-qubit state.}
\end{figure}
The toric polytope of  $\mathbf{X}_{\Sigma}=\mathbf{CP}^{1}\times\mathbf{CP}^{1}\times\mathbf{CP}^{1}$ is a 3-cube.
Now, if
we consider the same holonomic controlled quantum gate
$U_{CPhase}$
acting on a pure three-qubit product state
$\ket{\Psi}=H\otimes H\otimes H\ket{000}=\frac{1}{\sqrt{2^{3}}}(\ket{000}+
\ket{001}+\ket{010}+\ket{011}+\ket{100} +\ket{101}+\ket{110}+\ket{111}),$
 then we get
\begin{eqnarray}
(U_{CPhase}\otimes I)\ket{\Psi}&=&
\frac{1}{\sqrt{2^{3}}}(e^{i\varphi_{1}}( \ket{000}+\ket{001}+
\ket{110}+\ket{111}
\\\nonumber&+&
e^{i\varphi_{2}} (\ket{010}+\ket{100}+\ket{011}+\ket{101}))
\end{eqnarray}
where $I$ is 2-by-2 identity matrix. Now, we can see directly that phase factors $e^{i\varphi_{1}}$ and $e^{i\varphi_{2}}$
 correspond to intersecting diagonal planes in the toric variety of a given three-qubit state.
Moreover, we have
\begin{eqnarray}
(I\otimes U_{CPhase})\ket{\Psi}&=&\frac{1}{2^{3}}(e^{i\varphi_{1}}( \ket{000}+\ket{011}+
\ket{100}+\ket{111}),
\\\nonumber&+&
e^{i\varphi_{2}} (\ket{011}+\ket{010}+\ket{101}+\ket{110})).
\end{eqnarray}
In this  case the phase factors $e^{i\varphi_{1}}$ and $e^{i\varphi_{2}}$
 correspond to intersecting diagonal planes in the toric variety of a  three-qubit state, see Figure 2.
\subsection{Four qubit systems}
Finally, we will discuss a four-qubit state $\ket{\Psi}=\sum^{1}_{x_{4},x_{3},x_{2},x_{1}=0}
\alpha_{x_{1}x_{2}x_{3}x_{4}} \ket{x_{1}x_{2}x_{3}x_{4}}
$. In this case we can also  shows that the toric variety $\mathbf{X}_{\Sigma}=\mathbf{CP}^{1}\times\mathbf{CP}^{1}\times\mathbf{CP}^{1}
\times\mathbf{CP}^{1}$ is a four hypercube following the same procedure.
\begin{figure}\label{Figure 3}
\begin{center}
\includegraphics[scale=0.40]{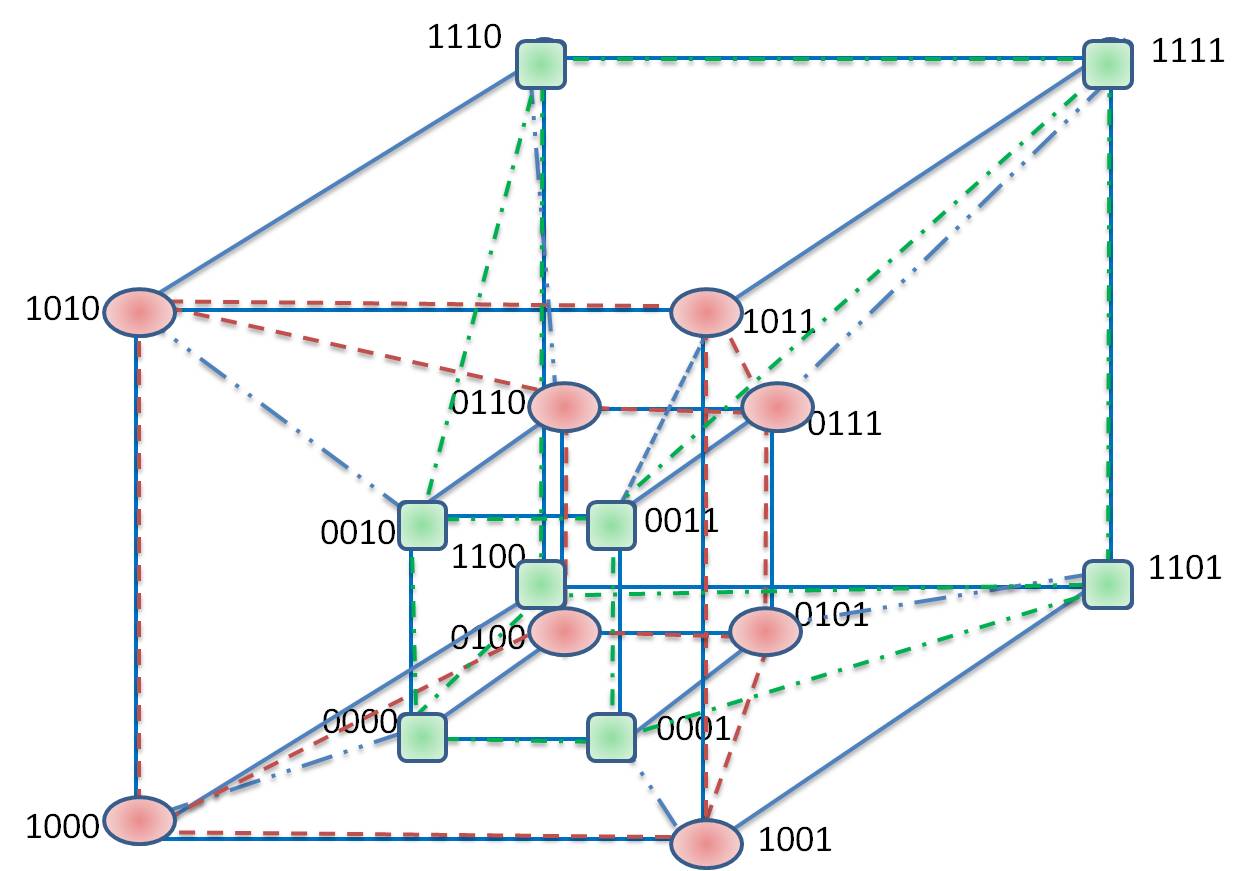}
\end{center}
\caption{Action of the holonomic controlled quantum gate on toric variety of a four-qubit state.}
\end{figure}
Now, we consider the same geometric quantum gate
$U_{CPhase}$
acting on a pure four-qubit product state
$\ket{\Psi}=(H\otimes H\otimes H\otimes H)\ket{0000}=\frac{1}{\sqrt{2^{4}}}(\ket{0000}+
\ket{0001}+\cdots+\ket{1111})
$, then we get
\begin{eqnarray}\nonumber
(U_{CPhase}\otimes I\otimes I)\ket{\Psi}&=&
\frac{1}{\sqrt{2^{4}}}(e^{i\varphi_{1}}( \ket{0000}+\ket{0001}+
\ket{0010}+\ket{0011}\\\nonumber&+&\ket{1100}+\ket{1101}+\ket{1110}+\ket{1111}
\\\nonumber&+&
e^{i\varphi_{2}} (\ket{0100}+\ket{0101}+\ket{0110}+\ket{0111}\\&+&\ket{1000}
+\ket{1001}+\ket{1010}+\ket{1011}))
\end{eqnarray}
Here we can also see that phase factors $e^{i\varphi_{1}}$ and $e^{i\varphi_{2}}$
 correspond to intersecting three cubes in the toric variety of a given four-qubit state.
We are not going to discuss the other possible cases $( I\otimes U_{CPhase}\otimes I)\ket{\Psi}$ and $( I\otimes I\otimes U_{CPhase})\ket{\Psi}$. But in these cases also we will have intersecting three cubes in side a 4-hypercube, see Figure 3.

In this paper we have visualized the action of a holonomic controlled quantum gate based on toric varieties. The result can be generalized into multi-qubit states. In example, the holonomic quantum gate defined by equation (3)  acts on a multi-qubit state  as a two intersecting $(m-1)$-hypercubes ($m\geq4$) in side a $m$-hypercube which is the toric variety of the system. There is also another interesting relation between toric varieties and  measures of entanglement for multi-qubits states such as $m$-tangle. And these combinatorial structures possibly have more applications in quantum information and quantum computing that need to be explored.


\begin{flushleft}
\textbf{Acknowledgments:}  The  work was supported  by the Swedish Research Council (VR).
\end{flushleft}

\end{document}